\RequirePackage{amsmath}

\documentclass[10pt,preprint,compress,twocolumn,3p]{elsarticle}

\usepackage{bm}
\usepackage{graphicx}
\usepackage{color}
\usepackage{amssymb}
\usepackage{bbold}
\usepackage{comment}

\DeclareMathOperator{\sgn}{sgn}
\renewcommand{\d}{\mathrm{d}}

\journal{Physica E}

\begin{document}

\begin{frontmatter}

\title{Tailoring topological states of core-shell nanoparticles}

\author[1]{Carolina\ Mart\'{\i}nez Strasser}
\author[1]{Yuriko Baba\corref{cor1}}
\author[2]{\'{A}lvaro D\'{\i}az-Fern\'{a}ndez}
\author[1]{Francisco Dom\'inguez-Adame}

\cortext[cor1]{Corresponding author}

\address[1]{GISC, Departamento de F\'{\i}sica de Materiales, Universidad Complutense, E-28040 Madrid, Spain}

\address[2]{GISC, Departamento de Estructuras y F\'{\i}sica de Edificaci\'{o}n, Universidad Polit\'{e}cnica de Madrid, E--28031 Madrid, Spain}

\begin{abstract}

In this work we investigate novel spherical core-shell nanoparticles with band inversion. The core and the embedding medium are normal semiconductors while the shell material is assumed to be a topological insulator. The envelope functions are found to satisfy a Dirac-like equation that can be solved in a closed form. The core-shell nanoparticle supports midgap bound states located at both interfaces due to band inversion. These states are robust since they are topologically protected. The energy spectrum presents mirror symmetry due to the chiral symmetry of the Dirac-like Hamiltonian. As a major result, we show that the thickness of the shell acts as an additional parameter for the fine tuning of the energy levels, which paves the way for electronics and optoelectronics applications.

\end{abstract}


\end{frontmatter}

\section{Introduction}   \label{sec:intro}

Band-inverted semiconductor heterostructures were already studied back in the 1980s and 1990s, in which the fundamental gap has opposite sign on each semiconductor. A remarkable feature is the existence of midgap interface states (see Refs.~\cite{Volkov1985,Volkov1987,Korenman1987,Agassi1988,Pankratov1990,Kolesnikov1997} and references therein). These states are protected by symmetry and are responsible for the conducting properties of the surface. Semiconductor materials that can present band inversion are II-VI compounds, such as {Hg}$_{x}${Cd}$_{1-x}${Te} and IV-VI compounds, such as {Pb}$_{x}${Sn}$_{1-x}${Te} and {Pb}$_{1-x}${Sn}$_{x}${Se}, among others. In this context, Dziawa \emph{et al.} provided strong evidence on how these IV-VI narrow-gap compounds become a topological insulator (TI) for $x=0.23$ at temperatures below the critical temperature, $T_{c}$, which is the temperature at which a transition between a normal state and an inverted bandgap state occurs for a Sn content of $0.18 \leq x \leq 0.3$~\cite{Dziawa2012}.

In recent years, nanoparticles based on TIs have been thoroughly investigated for their interest in photonics, optically controlled quantum memory and quantum computing~\cite{Paudel2013,Lin2015,Siroki2017,Siroki2016,Gioia2019,Castro2020,Chatzidakis2020}. Surface states show little sensitivity to disorder, which is beneficial for optical applications at the nanoscale~\cite{Siroki2017}. Siroki \emph{et al.} have found that, under the influence of light, a single electron in a topologically protected surface state creates a surface charge density, similar to a plasmon in a metallic nanoparticle~\cite{Siroki2016}. Hybrid systems composed of a topological-insulator nanoparticle and a quantum emitter dimer, interacting in the strong-coupling regime, show the emergence of a mode that stems from the coupling of the surface topological particle polariton of the topological insulator with the resonance state of the quantum emitter~\cite{Chatzidakis2020}.

By contrast, core-shell nanoparticles based on TIs have received much less attention. Yue \emph{et al.} reported on a novel conic plasmonic nanostructure that is made of bulk TIs and has an intrinsic core-shell formation. Through integration of the nanocone arrays into a-Si thin film solar cells, up to $15\%$ enhancement of light absorption was predicted in the ultraviolet and visible ranges~\cite{Yue2016}. In this paper we consider a novel spherical core-shell nanoparticle, where the core is made of a normal insulator (NI) and the shell is made of a TI. The nanoparticle is embedded in the same NI of the core. Consequently, the core-shell nanoparticle present two band-inverted heterojunctions, leading to the hybridization of the surface states when the width of the shell is of the order of the spatial decay length of the states. As a major result, we show that the thickness of the shell can be used for the fine tuning of the surface energy levels and hence the electric response of the system.

\section{Model Hamiltonian}   \label{sec:model}

The two-band model is a reliable approach to obtaining the electron states near the band edges in narrow-gap IV-VI semiconductors, for which the coupling to other bands is negligible~\cite{Agassi1988,Melngailis1972,Burkhard1979,Assaf2016}.  The electron wave function is written as a sum of products of band-edge Bloch functions with slowly varying envelope functions. The corresponding envelope function ${\bm\chi}({\bm r})$ is a four-component column vector composed by the two-component spinors ${\bm\chi}_{+}({\bm r})$ and ${\bm\chi}_{-}({\bm r})$ belonging to the two bands. Electron states near the band edges are determined from the Dirac-like equation $\mathcal{H}{\bm\chi}({\bm r})=E{\bm\chi}({\bm r})$ with~\cite{Agassi1988,Pankratov1990,Shen2012} 
\begin{equation}
\mathcal{H}=\hbar v\, {\bm\alpha}\cdot{\bm k} +\frac{1}{2}\,E_{\mathrm{G}}({\bm r})\beta+V_{\mathrm C}({\bm r})\ .
\label{eq01}
\end{equation}
Here $v$ is an interband matrix element having dimensions of velocity, $E_{\mathrm{G}}({\bm r})$ denotes the position-dependent gap and $V_{\mathrm C}({\bm r})$ gives the position of the gap center. ${\bm\alpha}=(\alpha_x,\alpha_y,\alpha_z)$ and $\beta$ denote the usual $4\times 4$ Dirac matrices
\begin{equation*}
\alpha_i=\begin{pmatrix}
\mathbb{0}_2 & \sigma_i \\
\sigma_i & \mathbb{0}_2
\end{pmatrix} \ ,
\quad
\beta=\begin{pmatrix}
\mathbb{1}_2 & \mathbb{0}_2 \\
\mathbb{0}_2 & -\mathbb{1}_2
\end{pmatrix} \ ,
\end{equation*}
$\sigma_i$ being the Pauli matrices ($i=x,y,z$), and $\mathbb{1}_n$ and $\mathbb{0}_n$ are the $n\times n$ identity and null matrices, respectively. In order to keep the algebra as simple as possible, we restrict ourselves to the symmetric situation with same-sized and aligned gaps [$V_\mathrm{C}({\bm r})=0$]. This is not a serious limitation, but the calculations are largely simplified. 

In the case of spherically symmetric nano\-particles, the gap function $E_{\mathrm{G}}(r)$ depends only on the distance $r=|{\bm r}|$ to the origin. This prompts us to introduce spherical coordinates by a separable ansatz
\begin{equation}
{\bm\chi}({\bm r})=\frac{1}{r}
  \begin{pmatrix}
  u_1(r)\,\mathcal{Q}_{\kappa,\mu}(\theta,\phi) \\
  iu_2(r)\,\mathcal{Q}_{-\kappa,\mu}(\theta,\phi) \\
  \end{pmatrix}\ ,
\label{eq02}
\end{equation}
where $u_1(r)$ and $u_2(r)$ take into account the radial part and $\mathcal{Q}_{\kappa,\mu}(\theta,\phi)$ are the eigenfunctions of the angular dependent part, as defined in Ref.~\cite{Greiner2000}:
\begin{align}
    \left( {\bm \sigma}\cdot {\bm L} +\hbar \right)\mathcal{Q}_{\pm\kappa,\mu}&=\mp \hbar \kappa \mathcal{Q}_{\pm\kappa,\mu}\ ,\nonumber \\
    J_z \mathcal{Q}_{\kappa,\mu} & =\hbar\mu\mathcal{Q}_{\kappa,\mu}\ .
\label{eq03}
\end{align}
Here the total angular momentum is ${\bm J}={\bm L}+{\bm S}$, $\kappa=\pm (j+1/2)=\pm 1, \pm 2, \pm 3,\ldots$, $\mu=-j,-j+1/2,\ldots, +j$, and $\ell = j \pm 1/2$. Inserting the ansatz~\eqref{eq02} into Eq.~\eqref{eq01} we obtain the following two coupled differential equations
\begin{subequations}
\begin{equation}
    \frac{\d\phantom{r}}{\d r} 
    {\bm U}(r)  
    =\begin{bmatrix}
    -\,\dfrac{\kappa}{r} & 
    \dfrac{\Delta(r)\!+\!E}{\hbar v} \\
    \dfrac{\Delta(r)\!-\!E}{\hbar v} &
    \dfrac{\kappa}{r} \\
    \end{bmatrix}\!
    {\bm U}(r)\ ,
\label{eq04}
\end{equation}
with $\Delta(r)=E_\mathrm{G}(r)/2$ being half of the position-dependent gap and
\begin{equation}
{\bm U}(r)=\begin{pmatrix}
            u_1(r) \\
            u_2(r) \\
           \end{pmatrix} \ .
\end{equation}
\end{subequations}

\section{Spherical quantum dot}   \label{sec:qd}

\begin{figure}[htbp]
    \centering
    \includegraphics[width=\linewidth]{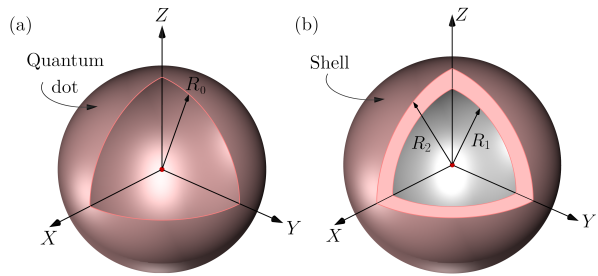}
    \caption{Schematic representation of (a)~a spherical TI quantum dot and (b)~a spherical core-shell nanoparticle, where the core (shell) is made of a TI (NI). In both cases the embedding medium is a NI.}
    \label{fig01}
\end{figure}

For the sake of completeness, we first consider a TI quantum dot of radius $R_{0}$ embedded in a NI, as shown in Fig.~\ref{fig01}(a) and addressed in Ref.~\cite{Paudel2013} using Green's function techniques associated with the squared Hamiltonian. Regarding the materials of choice, we take Pb$_{0.57}$Sn$_{0.43}$Te for the TI and Pb$_{0.67}$Sn$_{0.33}$Te for the embedding NI. With this choice of materials the magnitude of both gaps $\Delta$ are approximately the same and we can set $\Delta(r)\equiv \Delta\, s(r)$, where
\begin{equation}
    s(r)=\left\{
       \begin{array}{ll}
        -1\ ,    & 0\leq r \leq R_0\ ,  \\
        +1\ ,    & R_0 < r < \infty\ .
       \end{array}
    \right.
    \label{eq05}
\end{equation}

We now introduce the length scale $d=\hbar v/\Delta$, the dimensionless energy $\epsilon=E/\Delta$ and the dimensionless coordinate $\xi=r/d$. In these units, Eq.~\eqref{eq04} reads
\begin{equation}
    \frac{\d\phantom{r}}{\d \xi} 
    {\bm U}(\xi)=
    \begin{bmatrix}
    -\kappa/\xi & 
    s(\xi)\!+\!\epsilon \\
    s(\xi)\!-\!\epsilon &
    \kappa/\xi \\
    \end{bmatrix}\!
    {\bm U}(\xi)\ .
\label{eq06}
\end{equation}
Here $s(\xi)=\sgn(\xi-\xi_0)$ with $\xi_0=R_0/d$. Since the function $s(\xi)$ is piecewise constant, we can easily solve Eq.~\eqref{eq06} for $\xi<\xi_0$ (region~I hereafter) and $\xi>\xi_0$ (region~II hereafter). Our interest concerns midgap states ($|E|<\Delta$) and we then define the real parameter $\lambda=\sqrt{1-\epsilon^2}$. The solution to Eq.~\eqref{eq06} reads
\begin{align}
    u_1(\xi)&=\sqrt{\xi}\Big[A_iI_{|\kappa+1/2|}(\lambda\xi)\nonumber \\
    &+B_iK_{|\kappa+1/2|}(\lambda\xi)\Big]\ ,\nonumber\\
    u_2(\xi)&=\frac{s(\xi)-\epsilon}{\lambda}\sqrt{\xi}\Big[A_iI_{|\kappa-1/2|}(\lambda\xi)\nonumber \\
    & -B_iK_{|\kappa-1/2|}(\lambda\xi)\Big]\ ,
    \label{eq07}
\end{align}
where $i=\mathrm{I},\mathrm{II}$, $A_i$ and $B_i$ are integration constants, and $I_{\nu}(z)$ and $K_{\nu}(z)$ stand for the modified Bessel functions~\cite{Abramowitz1972}. For small arguments ($z\to 0$) we have~\cite{Abramowitz1972} 
\begin{subequations}
\begin{align}
    I_{\nu}(z)& \sim \left(\frac{z}{2}\right)^{\nu}\,\frac{1}{\Gamma(\nu+1)}\ , \quad \nu\neq -1,-2,\ldots\nonumber\\
    K_{\nu}(z)&\sim \left(\frac{z}{2}\right)^{-\nu}\,\frac{1}{2\Gamma(\nu)} \ ,\quad \nu\neq 0\ .
    \label{eq08a}
\end{align}
Consequently, we must set $B_\mathrm{I}=0$ to obtain regular solutions at the origin. Similarly, the asymptotic expansion for large arguments ($z\to\infty $) reads~\cite{Abramowitz1972}
\begin{align}
    I_{\nu}(z) & \sim \frac{1}{\sqrt{2\pi z}}\,e^z \left(1-\frac{4\nu^2-1}{8z}\right)\ , \nonumber\\
    K_{\nu}(z) & \sim  \sqrt{\frac{\pi}{2 z}}\,e^{-z} \left(1+\frac{4\nu^2-1}{8z}\right) \ .
    \label{eq08b}
\end{align}
\label{eq08}%
\end{subequations}
Therefore, normalizability of the envelope function requires $A_\mathrm{II}=0$. Finally, the continuity of $u_1(\xi)$ and $u_2(\xi)$ at the interface $\xi=\xi_0$ between the TI and the NI leads to
\begin{align}
   & A_\mathrm{I}I_{|\kappa+1/2|}(\lambda\xi_0)=B_\mathrm{II}K_{|\kappa+1/2|}(\lambda\xi_0)\ ,\nonumber\\
   & A_\mathrm{I}(1+\epsilon)I_{|\kappa-1/2|}(\lambda\xi_0)\nonumber \\
   &=B_\mathrm{II}(1-\epsilon)K_{|\kappa-1/2|}(\lambda\xi_0)\ ,
    \label{eq09}
\end{align}
whence
\begin{subequations}
\begin{equation}
    \mathcal{F}(\epsilon,\kappa,\xi_0)=0\ ,
    \label{eq10a}
\end{equation}
with
\begin{align}
    &\mathcal{F}(\epsilon,\kappa,\xi) \equiv (1+\epsilon)I_{|\kappa-1/2|}(\lambda\xi)K_{|\kappa+1/2|}(\lambda\xi) \nonumber\\
    & -(1-\epsilon)I_{|\kappa+1/2|}(\lambda\xi)K_{|\kappa-1/2|}(\lambda\xi)\ .
    \label{eq10b}
\end{align}
\label{eq10}
\end{subequations}
Notice that Eq.~\eqref{eq10a} is invariant under the change $\epsilon\to -\epsilon$ and $\kappa\to -\kappa$.

It is instructive to consider two limiting cases. In the first place, when $R_0\to \infty$ (i.e. $\xi_0\to \infty$), we might recover the energy level of a flat interface. This is indeed the case because in this limit we can take the asymptotic expansion~\eqref{eq08b} in Eq.~\eqref{eq10}, yielding $\epsilon\sim -\kappa/4\xi_0$. Undoing the change of variables we get $E\sim -\kappa \hbar v/4R_0$ and the energy of the topological interfaces states approaches the center of the gap on increasing $R_{0}$, as expected~\cite{Volkov1985,Volkov1987,Korenman1987,Agassi1988,Pankratov1990,Kolesnikov1997}. In the second place, we realize that there exists a minimum radius $R_\mathrm{min}$ for the quantum dot to support topological interface states. On decreasing the radius of the quantum dot $R_{0}$ from infinite (flat interface) to $R_\mathrm{min}$ the energy of the interface states moves away from the center of the gap ($E=0$) towards the band edges ($|E|=\Delta$). Hence, the condition to obtain $R_\mathrm{min}$ is $\lambda=0$ and we can then use the approximation~\eqref{eq08a} for small arguments in Eq.~\eqref{eq10}, yielding
\begin{equation}
    R_\mathrm{min}(\kappa)=\sqrt{\kappa^2-1/4}\,d=\sqrt{j(j+1)}\,d\ .
    \label{eq11}
\end{equation}

\begin{figure}[htbp]
    \centering
    \includegraphics[width=0.9\linewidth]{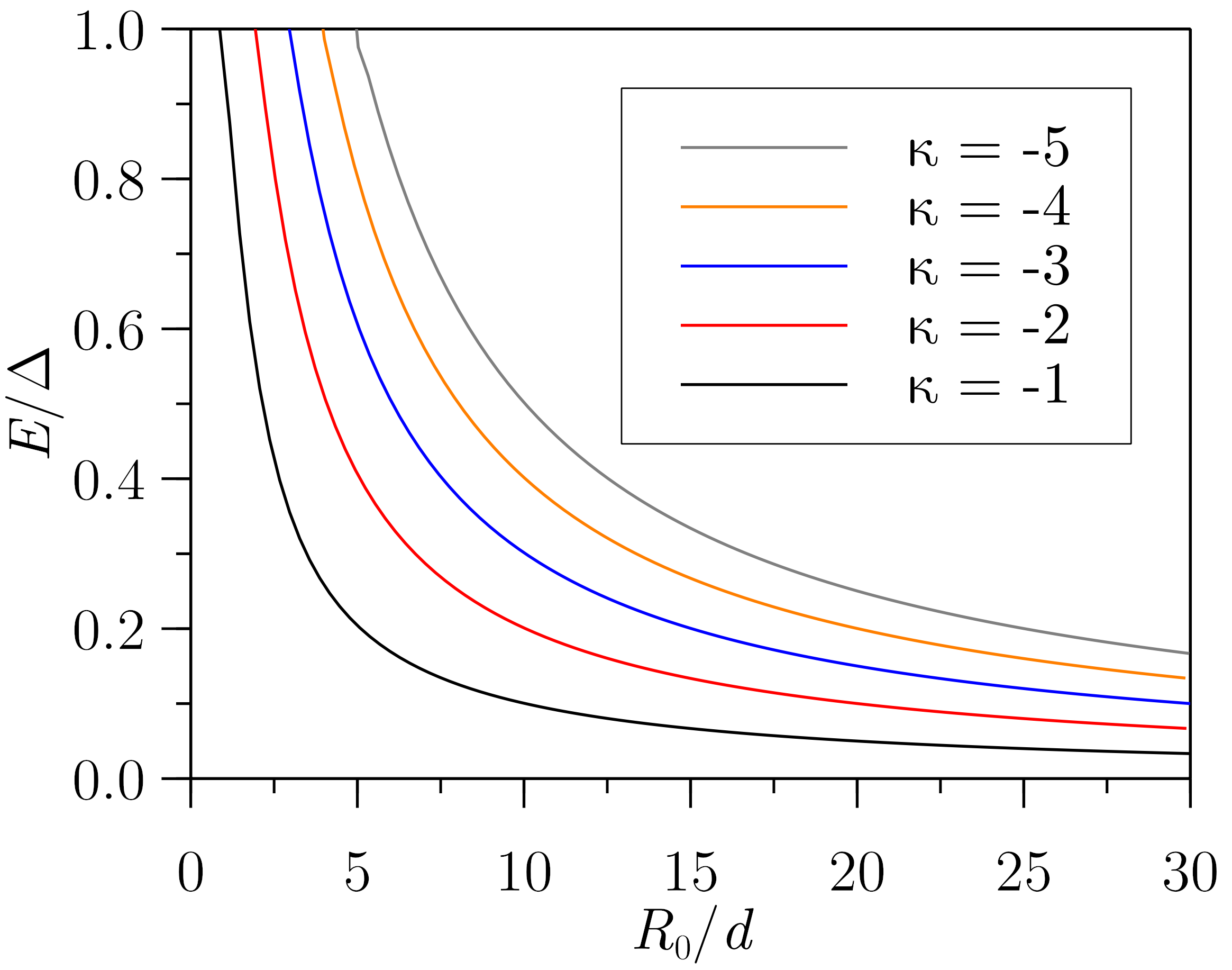}
    \caption{Energy of the topological interface states as a function of the radius of the quantum dot for $\kappa=-(j+1/2)<0$.}
    \label{fig02}
\end{figure}

Figure~\ref{fig02} shows the energy of the topological interfaces states as a function of the radius of the quantum dot for $\kappa<0$. We can obtain the energy levels for $\kappa>0$ reversing the sign of the energy due to the invariance mentioned above. As we already anticipated, the energy levels shift upwards from the gap center when $\kappa<0$ (and downwards for $\kappa>0$) on decreasing the radius of the quantum dot and reaches the band edge at $R_\mathrm{min}(\kappa)\simeq |\kappa|d$, according to Eq.~\eqref{eq11}.

\begin{figure}[htbp]
    \centering
    \includegraphics[width=\linewidth]{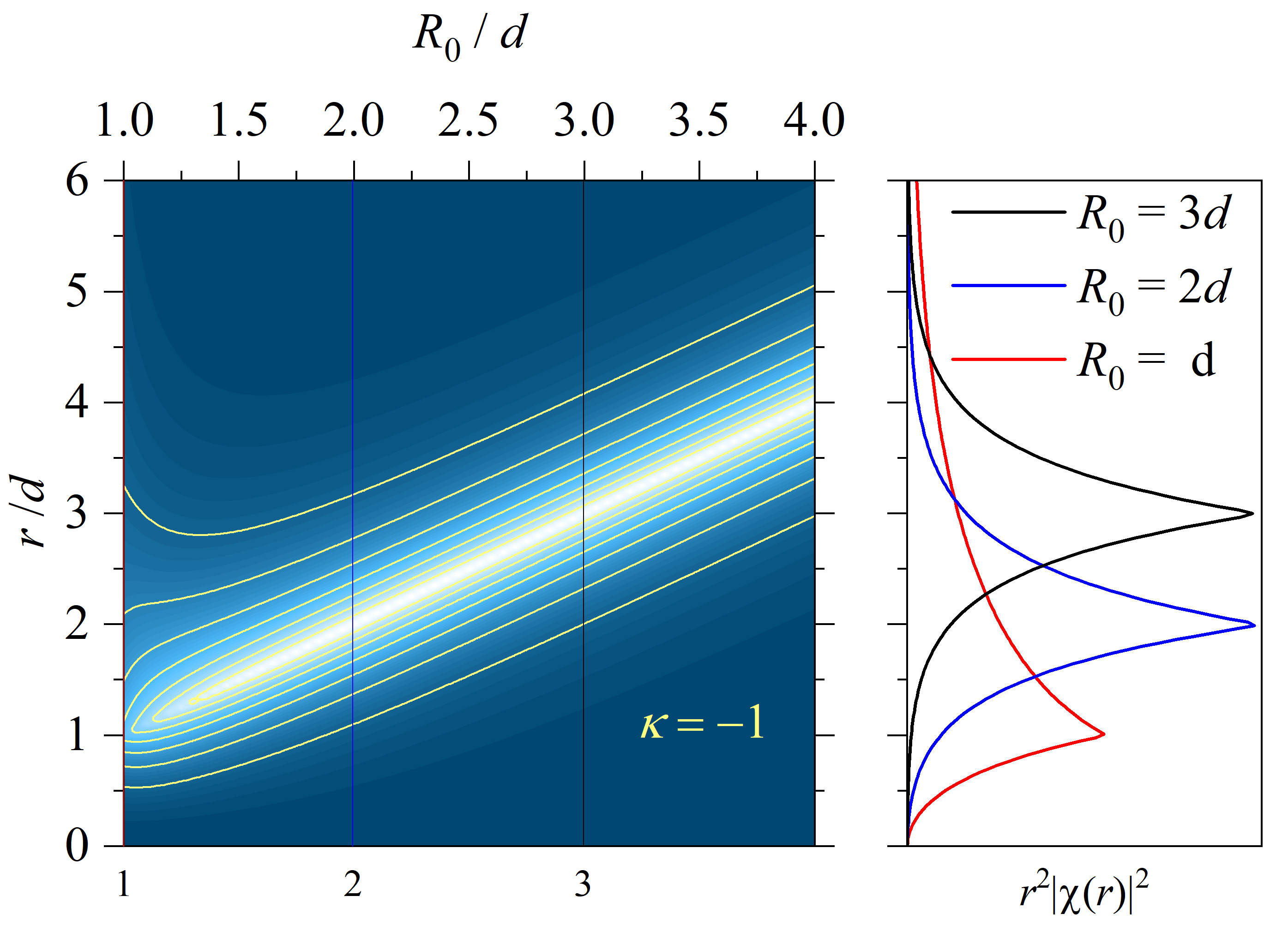}
    \caption{Left panel shows the contour plot of the radial probability density $P(r;R_0)$ of interfaces states with $\kappa=-1$ as a function of $r$ and $R_0$. Right panel shows the radial probability density of interfaces states at a function of $r$ for selected values of $R_0$, indicated in the legend, and corresponding to the vertical lines in the left panel. 
    }
    \label{fig03}
\end{figure}

The radial probability density of interfaces states, $P(r;R_0)\equiv r^2{\bm \chi}^{\dag}(r)\cdot{\bm \chi}(r)=u_1^2(r)+u_2^{2}(r)$ for $\kappa=-1$ is shown in Fig.~\ref{fig03}. 
When $R_0 \lesssim R_\mathrm{min}(\kappa)\simeq |\kappa|d$, the radial probability density presents a long tail outside the quantum dot (black lines in Fig.~\ref{fig03}). However, the radial probability density becomes almost symmetric at the interface as soon as $R_{0}$ exceeds $R_\mathrm{min}(\kappa)$, approaching the results for a flat interface.

\section{Spherical core-shell nanoparticle}   \label{sec:cs}

In this section we focus on the core-shell nanoparticle shown schematically in Fig.~\ref{fig01}(b). The inner (core) and embedding media are assumed to be the same NI, such as Pb$_{0.67}$Sn$_{0.33}$Te, while the shell is a TI, such as Pb$_{0.57}$Sn$_{0.43}$Te. The inner and outer radii are $R_1$ and $R_2$, respectively [see Fig.~\ref{fig01}(b)]. Thus, the gap profile in the nanoparticle is given as $\Delta(r) \equiv \Delta\, s(r)$, where now
\begin{equation}
    s(r)=\left\{
       \begin{array}{cl}
        +1\ ,    & 0 \leq r \leq R_1\ , \\
        -1\ ,    & R_1 < r <R_2\ ,  \\
        +1\ ,    & R_2 \leq r < \infty\ .
       \end{array}
    \right.
    \label{eq12}
\end{equation}
We introduce the same dimensionless magnitudes as before and define $\xi_1=R_1/d$ and $\xi_2=R_2/d$. The general solution to Eq.~\eqref{eq06} in the three regions of the nanoparticle, namely core (region~I), shell (region~II) and embedding medium (region~III), are still given by Eq.~\eqref{eq07} with $i=\mathrm{I},\mathrm{II},\mathrm{III}$. Normalizability of the envelope function implies that $B_\mathrm{I}=A_\mathrm{III}=0$. On the other hand, continuity of $u_1(\xi)$ and $u_2(\xi)$ at the interfaces $\xi=\xi_1$ and $\xi=\xi_2$ yields
\begin{align}
    A_\mathrm{I}&=A_\mathrm{II}+B_\mathrm{II}\,\frac{K_{|\kappa+1/2|}(\lambda\xi_1)}{I_{|\kappa+1/2|}(\lambda\xi_1)}\ , \nonumber\\
    A_\mathrm{I}&=-\,\frac{1+\epsilon}{1-\epsilon}\left[A_\mathrm{II}-B_\mathrm{II}\,\frac{K_{|\kappa-1/2|}(\lambda\xi_1)}{I_{|\kappa-1/2|}(\lambda\xi_1)}\right]\ , \nonumber\\
    B_\mathrm{III}&=B_\mathrm{II}+A_\mathrm{II}\,\frac{I_{|\kappa+1/2|}(\lambda\xi_2)}{K_{|\kappa+1/2|}(\lambda\xi_2)}\ , \nonumber\\
    B_\mathrm{III}&=-\,\frac{1+\epsilon}{1-\epsilon}\left[B_\mathrm{II}-A_\mathrm{II}\,\frac{I_{|\kappa-1/2|}(\lambda\xi_2)}{K_{|\kappa-1/2|}(\lambda\xi_2)}\right]\ .
    \label{eq13}
\end{align}
Setting the determinant to vanish we get
\begin{align}
    &\mathcal{F}(-\epsilon,\kappa,\xi_1)\mathcal{F}(\epsilon,\kappa,\xi_2) \nonumber \\
    &+ 4I_{|\kappa+1/2|}(\lambda\xi_1)I_{|\kappa-1/2|}(\lambda\xi_1)\nonumber \\
    &\times K_{|\kappa+1/2|}(\lambda\xi_2)K_{|\kappa-1/2|}(\lambda\xi_2)=0\ ,
    \label{eq14}
\end{align}
where $\mathcal{F}(\epsilon,\kappa,\xi)$ is defined in Eq.~\eqref{eq10b}. As in the case of the quantum dot, the symmetry  $\epsilon\to -\epsilon$ and $\kappa\to -\kappa$ is preserved as well. In the limit $\xi_1\to 0$, the second term of Eq.~\eqref{eq14} vanishes and $\mathcal{F}(-\epsilon,\kappa,\xi_1)$ remains finite. We then recover the results for a quantum dot of radius $\xi_2$, discussed in the previous section [see Eq.~\eqref{eq10a}].

\begin{figure}[htbp]
    \centering
    \includegraphics[width=0.9\linewidth]{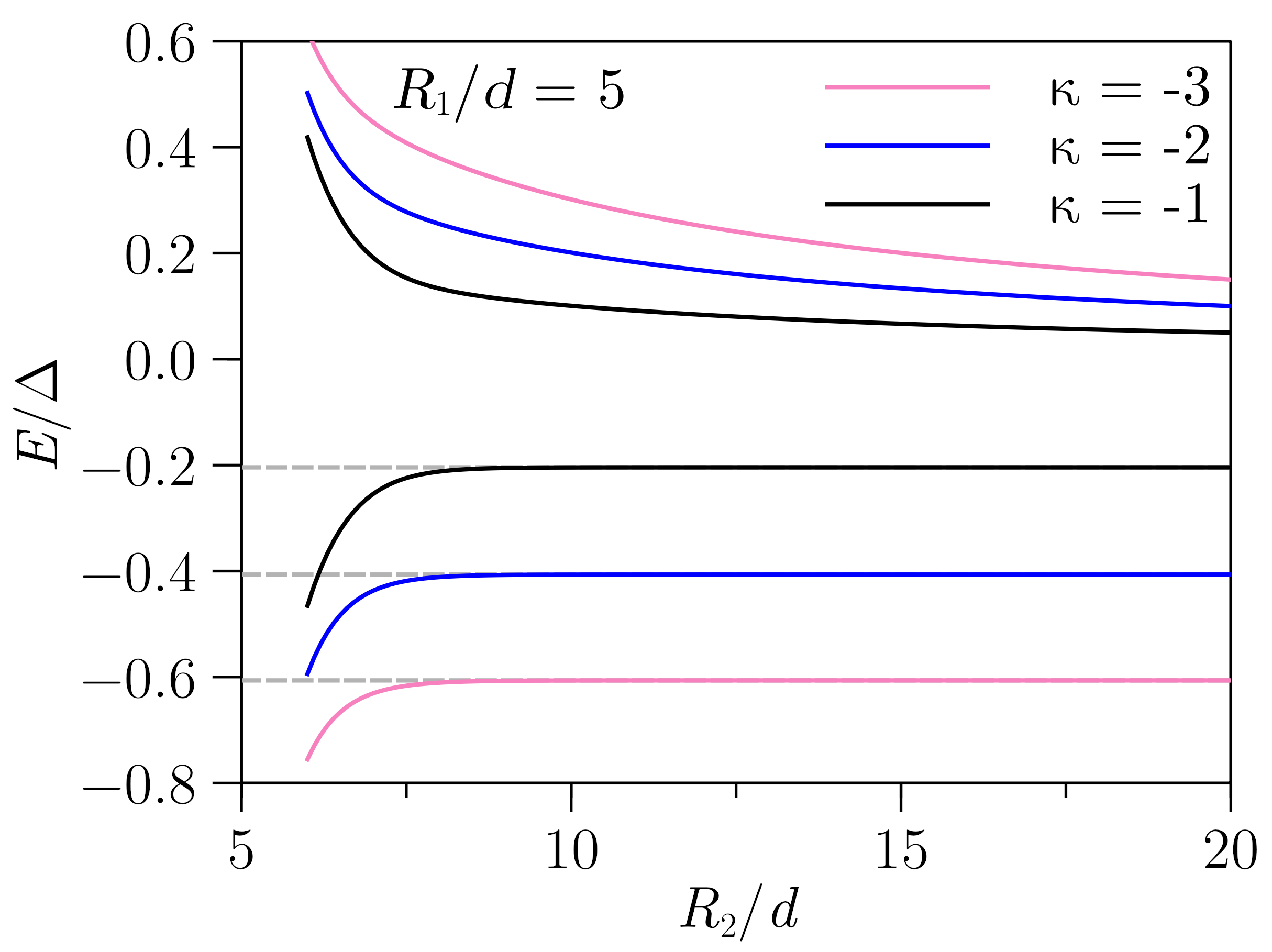}
    \caption{Energy of the topological interfaces states as a function of the outer radius of the core-shell nanoparticle for $\kappa=-(j+1/2)<0$. The inner radius is $R_1=5d$ and dashed lines indicate the energy levels of a quantum dot with radius $R_0=5d$.}
    \label{fig04}
\end{figure}

Figure~\ref{fig04} shows the energy of the topological interfaces states as a function of the outer radius $R_2$ for $\kappa<0$ and inner radius $R_1=5d$. The energy levels for $\kappa>0$ are obtained reversing the sign of the energy due to the symmetry mentioned above. Two set of states are observed. On the one side, energy levels above the center of the gap display the same behavior $\sim 1/R_2$ observed in quantum dots. Therefore, the corresponding states are mainly localized at the outer interface. On the other side, energy levels below the center of the gap approaches the energy levels corresponding to a quantum dot of radius $R_0=5d$ (see dashed lines in Fig.~\ref{fig04}). Thus, the corresponding states are localized at the inner interface. When $\kappa>0$, states above (below) are localised at the inner (outer) interface.

\begin{figure}[htbp]
    \centering
    \includegraphics[width=0.9\linewidth]{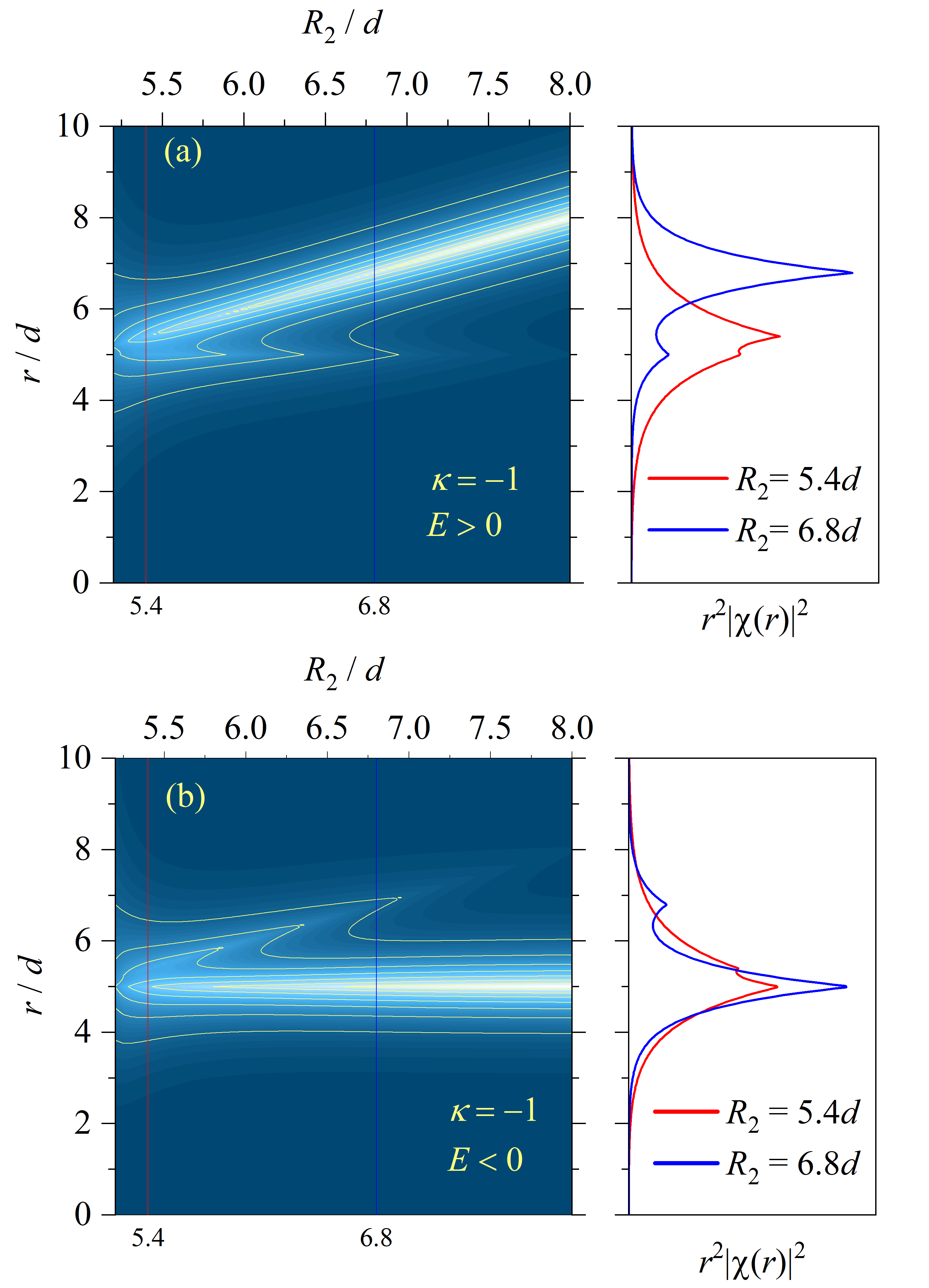}
    \caption{Left panels show the contour plot of the radial probability density $P(r;R_2)$ of interfaces states with $\kappa=-1$  and $R_1=5d$ as a function of $r$ for (a)~$E>0$ and (b)~$E<0$. Right panels show the radial probability density of interfaces states at a function of $r$ for selected values of $R_2$, indicated in the legend, and corresponding to the vertical lines in the left panel.}
    \label{fig05}
\end{figure}

Figure~\ref{fig05} corroborates the above statement regarding the spatial localization of the states. The radial probability density displays a small side peak at $R_1$ ($R_2$) for $E>0$ ($E<0$) in addition to the main peak at $R_2$ ($R_1$) when $R_2-R_1 \sim d$ (see blue lines in Fig.~\ref{fig05}). The small peak becomes unnoticeable when $R_2-R_1\gg d$ (see red lines in Fig.~\ref{fig05}).

It is then clear from the previous figure that the hybridization of the interface states can be greatly tuned by modifying the radii. Indeed, when $R_2-R_1\gg d$, the two interfaces are essentially decoupled, leading to the behaviour of two independent quantum dots, whereas they strongly couple when $R_2-R_1\sim d$. This is clearly observed in the fact that, as $R_2$ increases while keeping $R_1$ fixed, the smaller peak in the radial probability density vanishes, until a single peak at $R_1$ and $R_2$ remains for $E<0$ and $E>0$, respectively.

\section{Quadrupole moment of the core-shell nanoparticle}

Quantum dots and nanoparticles find a niche of applications in lasers, sensors, solar cells and single-electron transistors (see Ref.~\cite{Tong2020} and references therein). In the latter, Coulomb blockade plays a major role in the electrical response. Therefore, a detailed description of the electron-electron interaction is essential to properly explain the response of the device. The multipole expansion of the electrostatic potential created by a single electron inside the core-shell nanoparticle provides a direct way to determine the magnitude of the electron-electron interaction. Since the dipole field vanishes, we now focus on the quadrupole field created by an electron occupying a topological state of the core-shell nanoparticle. The components of the quadrupole tensor are given as
\begin{equation}
    Q_{ij}=-e\int \mathrm{d}^3{\bm r}\,|{\bm \chi}({\bm r})|^2 \left(3 x_i x_j - r^2\delta_{ij}\right)  \ .
    \label{eq15}
\end{equation}
We restrict ourselves to the calculation of $Q_{zz}$ hereafter. Defining $Q_0\equiv -e d^2$ and using Eq.~\eqref{eq02} one gets
\begin{align}
    \frac{Q_{zz}}{Q_0}&=\int_{0}^{\infty}\mathrm{d}\xi\, \xi^2 \int \mathrm{d}\Omega 
    \Big[ 
       u_1^2(\xi)\mathcal{Q}_{\kappa,\mu}^{\dag}(\Omega)\cdot\mathcal{Q}_{\kappa,\mu}(\Omega)
       \nonumber \\
      &+ u_2^2(\xi)\mathcal{Q}_{-\kappa,\mu}^{\dag}(\Omega)\cdot\mathcal{Q}_{-\kappa,\mu}(\Omega)
    \Big]
    \left(3\cos^{2}\theta-1\right)\ ,
    \label{eq16}
\end{align}
where $\Omega$ is the solid angle. After lengthy but straightforward calculations, one can find 
\begin{align}
    \frac{Q_{zz}}{Q_0} &=\frac{j(j+1)-3\mu^2}{2j(j+1)}\nonumber \\
    &\times \int_{0}^{\infty}\mathrm{d}\xi\, \xi^2 \left[u_1^2(\xi)+u_1^2(\xi)\right]\ .
    \label{eq17}
\end{align}

\begin{figure}[htbp]
    \centering
    \includegraphics[width=0.9\linewidth]{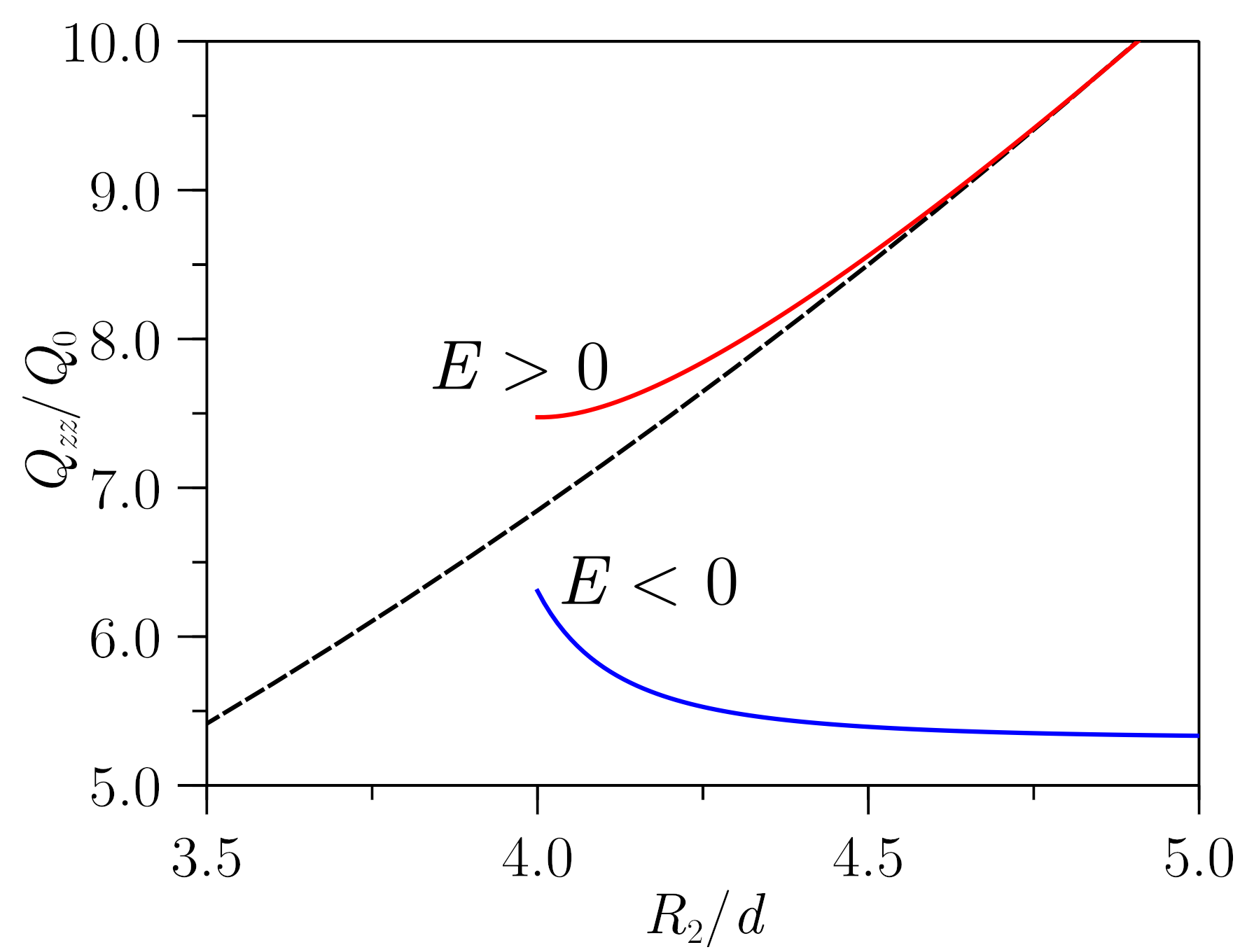}
    \caption{Quadrupole moment $Q_{zz}$ in units of $Q_0=-ed^2$ of the core-shell nanoparticle with a single electron in the state with $j=3/2$, $\mu=1/2$ and $\kappa=-(j+1/2)=-2$, as a function of $R_2$ when $R_1=3.5d$. Solid red and blue lines correspond to the state with positive and negative energy, respectively. For comparison, the black solid line shows the result for a quantum dot of the same radius.}
    \label{fig06}
\end{figure}

Figure~\ref{fig06} shows $Q_{zz}$ as a function of $R_2$ when $R_1=3.5d$ and the core-shell nanoparticle is occupied by a single electron in the state with $j=3/2$, $\mu=1/2$ and $\kappa=-(j+1/2)=-2$. It is worth mentioning that we obtain the same magnitude of the quadrupole moment but opposite sign when $\mu=3/2$. As we have already shown in Fig.~\ref{fig04}, for a given value of $\kappa$ there exist two states, one with positive energy and the other one with negative energy. In Fig.~\ref{fig06}, the solid red (blue) line shows the result for the positive (negative) energy state. For the state with positive energy, located at the outermost interface, the quadrupole moment approaches the value obtained in a quantum dot of the same radius as soon as $R_2$ exceeds $R_1+d$. On the contrary, for the state with negative energy, located in the innermost interface, the quadrupole moment is much lower and rapidly approaches the value corresponding a quantum dot of radius $R_0=3.5d$. This suggests that the single-electron state can be assessed by measuring the quadrupole moment.

\section{Conclusions}   \label{sec:conclusion}

Topological insulators at the nanoscale are  envisaged  to  have  an  ever-increasing number of applications. However, a more complete understanding of the properties of these materials is needed in order to better exploit these applications. In this work, we have proposed and studied a novel spherical core-shell nanoparticle system composed of a TI shell and a NI core and embedding medium. A realistic two-band description of the nanoparticle allowed us to describe topologically protected electron states that arise at the TI/NI interface. In addition, we were able to exactly solve the equation for the two-component envelope function. We found that there exists a strong hybridization of the topologically protected electron states located at the innermost and outermost interfaces, provided that the separation between the two surfaces is of the order of the decay length. A similar hybridization of topological states has been found in quantum wells~\cite{Diaz-Fernandez2017}. Since the hybridization can be controlled by selecting the thickness of the shell, we argue that it acts as an additional parameter for fine tuning the energy levels, which paves the way for electronics and optoelectronics applications.


\medskip

This work was supported by Minis\-terio de Cien\-cia e In\-novaci\'{o}n (Grant PID2019-106820RB-C21).

\section*{References}

\bibliography{references}

\bibliographystyle{model1a-num-names.bst}

\end{document}